\documentclass[12pt,preprint]{aastex}

\shorttitle{Kinematic and Chemical Analysis of LCSSPASGC  }
\shortauthors{Zhao et al.}

\begin{document}
\title{Two Portions of Sagittarius Stream in the LAMOST Complete Spectroscopic Survey of Pointing Area at Southern Galactic Cap}

\author{J. K. Zhao\altaffilmark{1}, X. H. Ye\altaffilmark{1,2}, H. Wu\altaffilmark{1,2}, M. Yang\altaffilmark{1,3}, Terry D. Oswalt\altaffilmark{4}, X. X. Xue\altaffilmark{1}, Y. Q., Chen\altaffilmark{1,2}, J.J. Zhang\altaffilmark{1,2}, G. Zhao\altaffilmark{1,2}   }

\altaffiltext{1}{CAS Key Laboratory of Optical Astronomy, National Astronomical Observatories,  Chinese Academy of Sciences£¬Beijing 100101, China. zjk@nao.cas.cn, myang@nao.cas.cn, hwu@nao.cas.cn }
\altaffiltext{2}{School of Astronomy and Space Science, University of Chinese Academy of Sciences, Beijing 100049, China}
\altaffiltext{3}{IAASARS, National Observatory of Athens, Vas. Pavlou and I. Metaxa, Penteli 15236, Greece }

\altaffiltext{4}{Embry-Riddle Aeronautical University, Aerospace Boulevard, Daytona Beach FL, USA, 32114. oswaltt1@erau.edu}


\begin{abstract}

We constructed a sample of 13,798 stars with $T\rm_{eff}$, log $g$, [Fe/H], radial velocity, proper motions and parallaxes from LAMOST DR5 and Gaia DR2 in the LAMOST Complete Spectroscopic Survey of Pointing Area (LaCoSSPAr) at the Southern Galactic Cap consisting of areas A and B. Using the distributions in both proper motions and radial velocity, we detected very significant overdensities in these two areas.  These substructures most likely are portions of  Sagittarius (Sgr) stream. With the Density-Based Spatial Clustering of Applications with Noise (DBSCAN) algorithm, 220 candidates stream members were identified.  Based upon distance to the Sun and published models, 106 of these stars are likely to be the members of the Sgr stream. The abundance pattern of these members using [$\alpha$/Fe] from Xiang et al. were found to be similar to Galactic field stars with [Fe/H] $<$ -1.5 and deficient to Milky Way populations at similar metallicities with [Fe/H] $>$ -1.0.  
 No vertical and only small radial gradients in metallicity along the orbit of Sgr stream were found in our Sgr stream candidates.

\end{abstract}


\section{Introduction}

 The most prominent tidal stream around the Milky Way is believed to comprise of remants of the Sgr dwarf spheroidal galaxy (dSph, Ibata et al. 1994, 2001; Majewski et al. 2003), which has been traced entirely around the Milky Way (Majewski et al. 2003; Sesar et al. 2017).  A wide variety of stellar spectral types have been used to trace the Sgr tidal debris such as main-sequence turn-off stars, blue horizontal branch (BHB) stars, red giants, RR Lyrae, and M giants.

 The orbit of Sgr stream has been well understood; two models (Law $\&$ Majewski 2010, LM10; Dierickx $\&$ Loeb 2017) have been shown to reproduce the properties of the stream.  Recently, Antoja et al. (2020) presented a measurement of the proper motion of the Sgr stream using Gaia DR2 data. Their determination is based on a much larger sample of stars than previous studies and covers different stellar types. For the first time, they also demonstrated that the Sgr stream extends over 2$\pi$ in the sky, except for dense region near the Galactic plane.

Ibata et al. (2020) presented the first full six-dimensional panoramic portrait of the Sgr stream using Gaia DR2 data and the STREAMFINDER algorithm. Using the precision distances provided by Gaia RR lyrae stars, they found that the  global morphological and kinematic properties of the Sgr stream are reasonably  well reproduced by the LM10 although that model overestimated the distances by up to $\sim$15\%. Ramos et al. (2020)  characterized the trends along the stream in 5 astrometric dimensions plus metallicity, covering more than 2$\pi$ rad in the sky. They also obtained new estimates for the apocenters and the mean metallicity [Fe/H] of the RR Lyrae population. Except orbit, the chemical properties have also been investigated.  Monaco et al. (2007), Chou et al. (2010), and Keller et al. (2010) analyzed Sgr M-giant stars, and found that on average, they are more metal-poor than
typical Sgr core stars, and are deficient in $\alpha$-elements relative to MW populations at similar metallicities. Carlin et al. (2018) picked up 42 Sgr stream stars from LAMOST M-giants. Using high-resolution spectra, they found stars in trailing and leading streams have systematic differences in [Fe/H], and the
$\alpha$-abundance patterns of Sgr stream is similar to those observed in Sgr core and other dwarf galaxies like the
large Magellanic Cloud and the Fornax dSph.

The Sgr dSph is known to have a complex star formation history. Ibata et al. (1995) showed that it contains a strong intermediate-age population (  4 $ \preceq \tau \preceq$ 8 Gyr) and metallicity (-0.6$\preceq$ [Fe/H] $\preceq$ -0.2). Siegel et al. (2007) demonstrated that Sgr has experienced at least 4 - 5 star formation bursts, including an old population: 13 Gyr and [Fe/H] = -1.8 dex from main sequence (MS) and red-giant branch (RGB) stars; at least two intermediate-aged populaitons: 4 - 6 Gyr with [Fe/H]= -0.6 to -0.4 dex from RGB stars; a 2.3 Gyr population near solar abundance: [Fe/H] = -0.1 dex from main sequence turn-off (MSTO)
stars.

Hasselquist et al. (2019) used chemical tagging to identify 35 relatively metal-rich ([Fe/H] $\geq$ -1.2) Sgr stream stars in the the Apache Point Observatory Galactic Evolution Experiment (APOGEE, Majewski et al. 2017).  Hayes et al. (2020) identified 166 Sgr stream members observed by APOGEE that also have Gaia DR2 astrometry.  Combined with the
APOGEE sample of 710 members of the Sgr dwarf spheroidal core, they established
differences of 0.6 dex in median metallicity and 0.1 dex in [$\alpha$/Fe] between  the Sgr core and dynamically older stream
samples.

Sesar et al. (2017) reported the detection of
spatially distinct stellar density features near the apocenters of
the Sgr stream's main leading and trailing arm, and found a
`spur' extending to 130 kpc at the apocenter of the trailing
arm using Pan-STARRS1 Type ab RR Lyrae (RRab) stars.

A Key Project named the LAMOST Complete Spectroscopic Survey of Pointing Area (LaCoSSPAr)  at  the Southern Galactic Cap (SGC)  is  designed  to  make  repeated spectroscopic  observations  of  all  sources  (Galactic  and  extragalactic)  in  two  20  deg$^{2}$ fields (A and B) in the  SGC (Yang et al. 2018).  The  central  coordinates  of  the  fields  are ($\alpha, \delta$) = (37.88150939$\rm^{o}$, 3.43934500$\rm^{o}$) and (21.525988792$\rm^{o}$, -2.200949833$\rm^{o}$), respectively. Targets in this field mainly consist of stars , galaxies, $u$ - band variables and QSOs (Lam et al. 2015; Cao et al. 2016). The stars and galaxies were selected from SDSS Data Release Nine (DR9; York et al. 2000) using $r$-band PSF magnitudes and Petrosian magnitudes between 14.0 $<$ r $<$ 18.1 mag, respectively. Compared with the LAMOST normal survey (Cui et al. 2012; Zhao et al. 2006, 2012), LCSSPA has better completeness and a deeper limiting magnitude. Both area A and B are located in the spatial position of the Sgr stream as identified in the LM10 model. {\bf  Even if so much research on Sgr stream, there are still some details needed to be exploited for Sgr stream. To understand the chemical abundance pattern of this stream using more elements, brighter member stars are required to do high resolution spectra observation, especially for FGK stars. Thus, LaCoSSPAr provides a good opportunity to identify the brighter member stars of this stream. Also, metallicity and [$\alpha$/Fe] distribution along the orbits of Sgr could be analyzed.}

In Section 2 we describe the general properties of  areas A $\&$ B.  Section 3 discusses the detection of Sgr stream in these areas. The chemical property of the member stars in the stream are presented in Section 4. A summary of our results is given in Section 5.

\section{Data}


\begin{figure}
\epsscale{.80}
\plotone{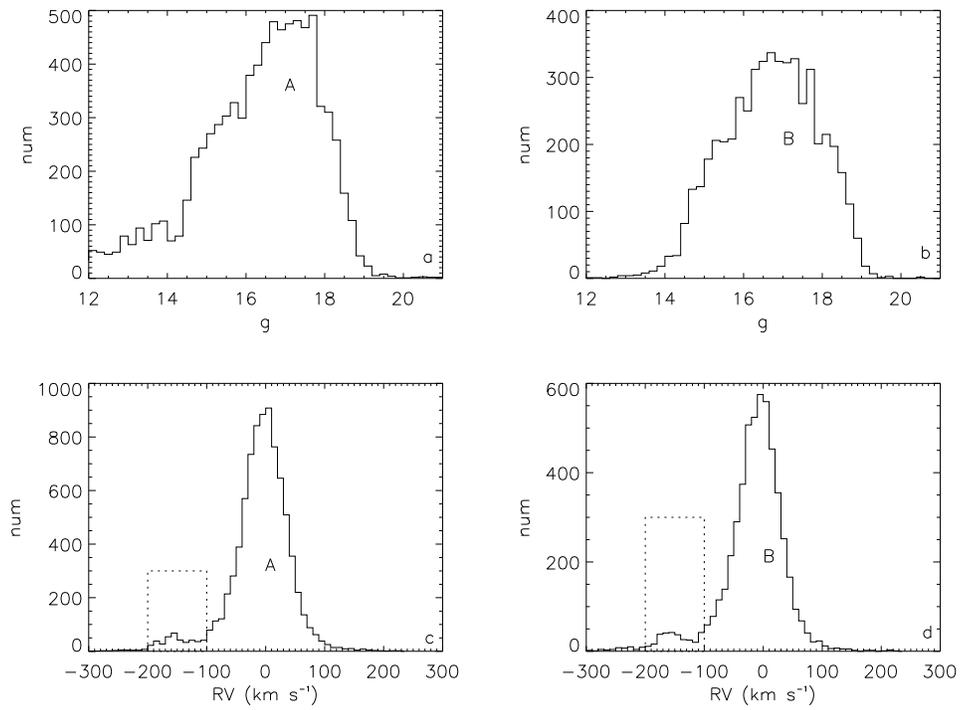}
\caption{Panel a:  $g$ magnitude distribution of stars in area A; panel b:  $g$ magnitude distribution of stars in area B; panel c, d: RV distribution of stars in areas A and B. The rectangles with dashed lines indicate the
overdensities in RV distribution.    \label{Fig1}}
\end{figure}

The LAMOST telescope is a 4 m Schmidt telescope at the Xinglong Observatory, National Astronomical Observatories of China (NAOC); this National Key Scientific
facility was built by the Chinese Academy of Sciences (Cui et al. 2012; Zhao et al. 2006, 2012). LAMOST finished its pilot survey in 2012 and the first-five-years regular
survey in 2017, respectively. The data from both surveys make up the fifth LAMOST data release (DR5), which contains optical band spectra (e.g., 3690 $-$ 9100$\rm \AA$) for 8,171,443 stars, 153,090 galaxies, 51,133 quasars, and 642,178 unknown
objects.

The total sample in two areas of LaCoSSPAr survey from the LAMOST DR5 includes 31,934 (A:17,694; B: 14,240) stars.
Most have been observed more than once. About 13,800 stars (A: 8,491; B:5,307) have stellar parameters from LAMOST pipeline LASP (Luo et al. 2015). For stars with $T\rm_{eff} <$ 8,000 K, temperature ($T\rm_{eff}$), gravity (log \textit{g}) and metallicity ([Fe/H]) are provided to precision (standard deviation) 110 K, 0.19 dex and 0.11 dex, respectively. Radial Velocity (RV) for stars cooler than 10,000 K are provided to a precision 4.91 km $\rm s^{-1}$.
Abundances [$\alpha$/Fe] from Xiang et al. (2019) have a typical precision  of 0.03 $<$ [$\alpha$/Fe] $<$ 0.1 dex, for spectra with signal to noise (SNR $>$ 50). By cross-matching with the Gaia DR2, proper motions and parallaies have been obtained for all the stars in the sample.

 Since Gaia parallaxes are not very accurate for distant stars beyond $\sim$ 4 kpc,  the distance of giants in our sample were estimated using a Bayesian method described in Xue et al. (2014),  which is based on the color magnitude diagrams (so-called fiducials) of three globular clusters and one open cluster observed by SDSS. First, we obtained the  photometries for our giants by cross-matching with Pan-STARRS1 (PS1; Chambers et al. 2016).  Then, PS1 magnitudes were converted to the SDSS system using
linear functions of $(g - i)_{P1}$ (Finkbeiner et al. 2016) for LAMOST K giants in common with both PS1
and SDSS (Xue et al. 2020, in preparation).
{\bf The E(B-V) was estimated from Schlegel et al. (1998).}  The median distance precision for this procedure is about 16\%. For each star, we calculated the line-of-sight Galactic
standard of rest velocity from the heliocentric RV using the formula:
\begin{equation}
V_{los}=RV+8.5coslcosb+244.5sinlcosb+13.38sinb
\end{equation}
This formula removes the contributions from the 238
km $\rm s^{-1}$ (Sch\"{o}nrich 2012) rotation velocity at the solar circle as
well as the solar peculiar motion (relative to the local standard
of rest) of $(U, V, W)_{0}$ = (8.5, 6.49, 13.38) km $\rm s^{-1}$ (Co\c{s}kuno\v{g}lu et al. 2011).
The Cartesian velocity components relative to the Sun for these stars
were then computed and transformed into Galactic velocity components \textit{U}, \textit{V}, and \textit{W}, and
corrected for the peculiar solar motion. The \textit{UVW}-velocity components are defined as a left-handed system with
\textit{U} positive in the direction radially outward from the Galactic center, \textit{V} positive in the
direction of Galactic rotation, and \textit{W} positive perpendicular to the plane of the Galaxy in the direction of the north Galactic pole.

The top panels of Fig. 1 show the $g$  magnitude distribution of our sample in two areas A $\&$ B. The \textit{g} magnitudes ranged from 12 to 19.5. It is clear that the number of stars of \textit{g} magnitude increases from  14 to 17.8, which indicates the completeness of these two areas is better than the areas in the normal survey that the star number decreases from \textit{g} $\sim$ 16.5. The RV distributions are presented in bottom panels of Fig. 1. Two peaks are apparent in the RV distributions of both areas A and B between -200 km s$\rm ^{-1}$ to -100 km s$\rm ^{-1}$, indicated by rectangles with dotted line. We speculate the bumps might be connected with substructures of the Milky Way.

\section{Testing substructures in areas A and B}
The bimodal RV distribution promoted us to look for similar substructures by proper motions and $V_{los}$.
\begin{figure}
\epsscale{.80}
\plotone{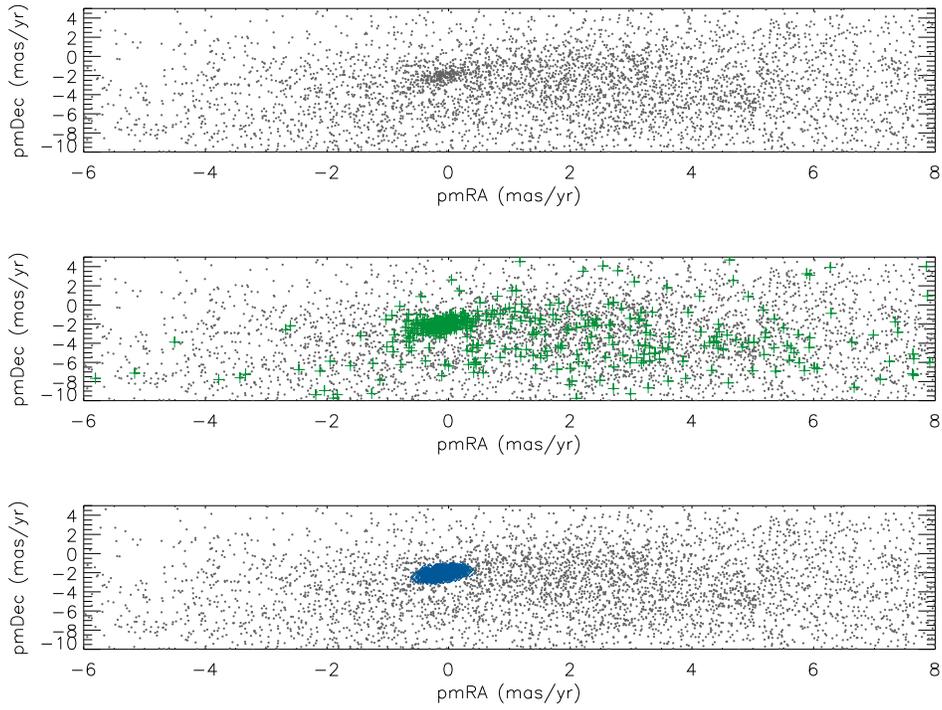}
\caption{ Proper motion distribution of area A. Top panel: proper motion distribution for all the stars in area A. Middle panel: gray points represent all the stars in area A while green pluses represent giants in Area A. Bottom panel: gray points are stars in area A. Blue diamonds are member candidates in the overdensity with DBSCAN algorithm.    \label{Fig2}}
\end{figure}

\subsection{The proper motion distributions}
Fig. 2 displays the procedure by which we identified substructure using proper motion in area A. The same procedure was used for area B. The top panel shows the proper motion distribution of stars in area A.  There is an overdensity centered at the (0, -2) mas $\rm yr^{-1}$ position. Considering the streams are distant and most members are giants, we divided the sample into dwarfs and giants{\bf . Stars with log $g < $ 3.5 were regarded as giants and others were assumed to be dwarfs}. In the middle panel, the green pluses are giants. Evidently the overdensity is more prominent in the giants sample. Thus, we conclude this overdensity is mostly composed of evolved stars, which indicates they are relative distant. To determine the most likely  member candidates in this substructure, we adopted the DBSCAN algorithm(Easter et al. 1996),  a well-known clustering algorithm, to identify cluster stars by $\left( \mu_{\alpha}, \mu_{\delta} \right)$. DBSCAN is based on the density and has no particular bias that depends on the shape of a distribution of points. In this work, we setup the parameters $eps = 0.20$ and $minPts = 30$ manually and identified 173 sources in the area A. In area B we set $eps = 0.30$ and $minPts = 25$ and identified 128 stars.

\begin{figure}
\epsscale{1.0}
\plottwo{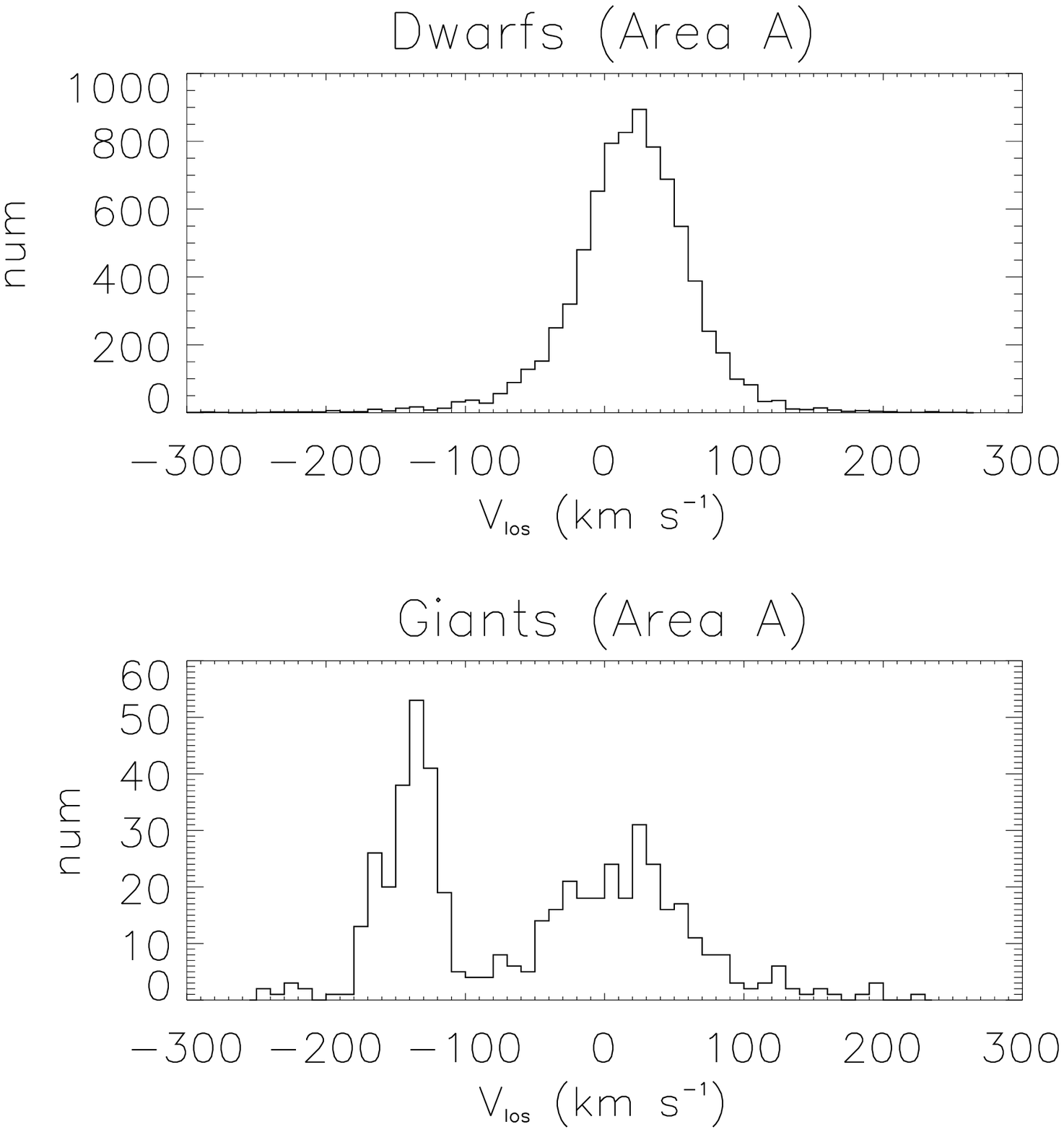}{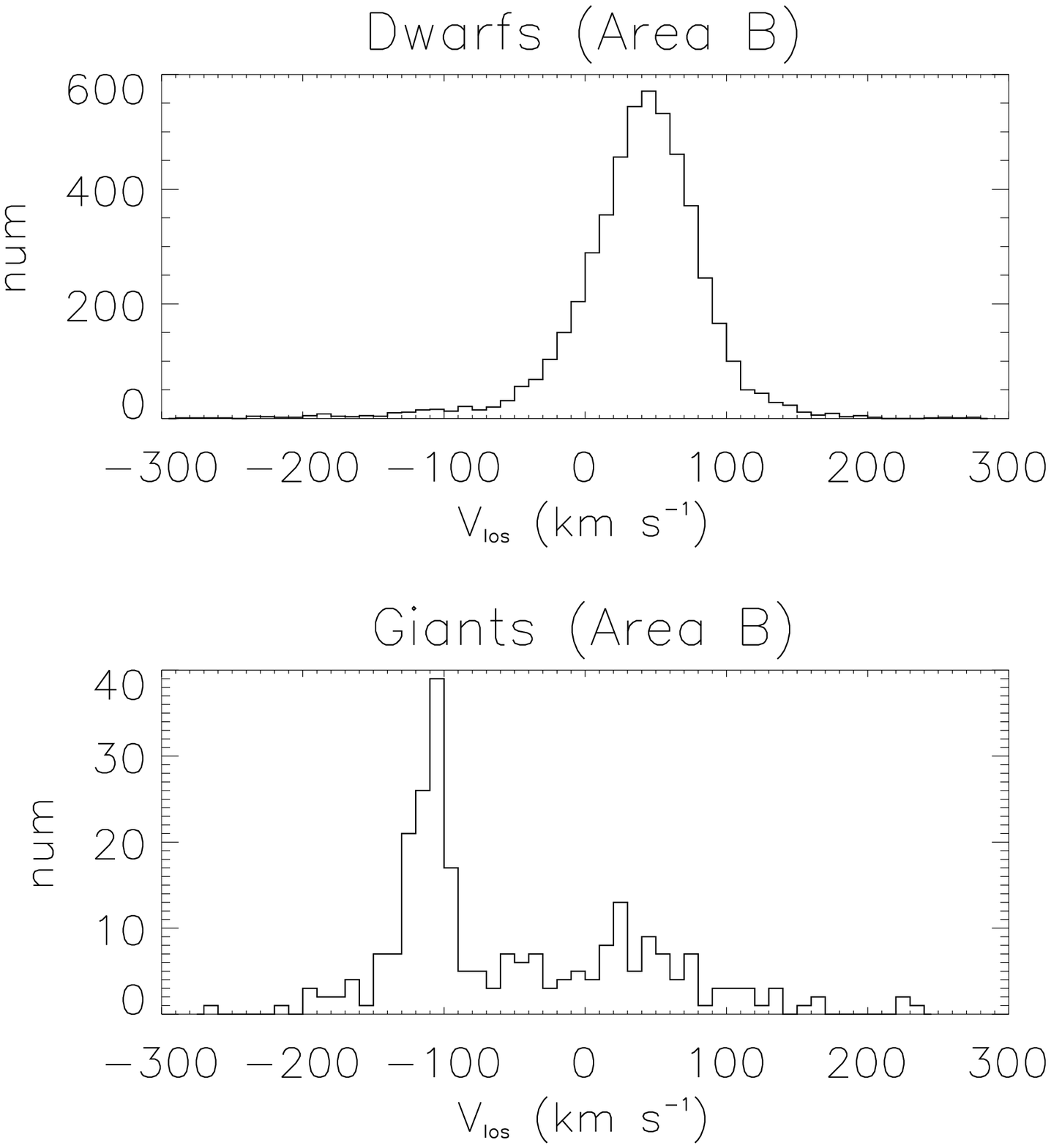}
\caption{ Overdensities in $V\rm_{los}$ distributions of areas A and B. Top panels: $V\rm_{los}$ distribution for {\bf dwarfs} of area A and B. bottom panel: $V\rm_{los}$ distribution  for giants. \label{Fig3}}
\end{figure}

\begin{figure}
\epsscale{1.0}
\caption{The spatial position, proper motion,  $V\rm_{los}$ and distance to the Sun distribution in the trailing arm (left panels) and leading arm (right panels) of Sgr stream from LM10 model. Pink points are the member candidates in areas A and B. \label{Fig4}}
\end{figure}

\subsection{Overdensities in $V\rm_{los}$ distribution}
 Fig. 3 shows the $V\rm_{los}$ distribution of stars in areas A and B. Top panels and bottom panels show the $V\rm_{los}$ distributions for both dwarfs and giants and giants, respectively. In the bottom panel, we find very prominent peaks in the $V\rm_{los}$ distribution for giants in both areas A and B.

\subsection{Identification of  member candidates}
The member candidates were initially identified by both proper motions and $V\rm_{los}$. For proper motions, the member candidates selected by DBSCAN algorithm as dwarfs were removed because they would be two nearby to be members of streams. Only giants with $V\rm_{los}$ ranging in [-180, -110] km s$^{-1}$ for area A and [-150, -90] km s$^{-1}$  for area B are considered as candidates. Thus, with restrictions from the proper motions and $V\rm_{los}$,  220 stars as initial member candidates were determined. The distances implied by Gaia parallaxes for these 220 stars are all less than 10 kpc indicating they are not accurate for distant star. Thus, we adopted the photometries distance for these giants.
{\bf The typical uncertainties of $T\rm_{eff}$, log $g$, [Fe/H] and [$\alpha$/Fe] of these member candidates are 120 K, 0.4 dex, 0.25 dex, 0.13 dex, respectively.  }

\subsection{Connection with known substructures}
 Using the high precision photometry $\&$ spectroscopy digital sky survey such as SDSS, APOGEE, RAVE, etc. have identified a number of substructures in the Milky Way.  Except Sgr stream, more substructures have been detected, such as orphan stream, monoceros ring, Hercules-Acuila Cloud, Virgo Overdensity etc.  The LaCoSSPAr overlaps with the Sgr stream fields, then we tentatively identified the overdensities detected above with this stream.

 Fig. 4 illustrates  the possible connection between the Sgr stream and the substructures in areas A and B using: spatial positions (panels a-b), proper motions(panels c-d), $V\rm_{los}$ distribution (panels e-f) and distance to the Sun (panels g-h). The black points are particles in the LM10 model for the Sgr stream. The pink points are our stream candidates in two substructures.   The left panels correspond the trailing arm of the Sgr stream, while the right panels shows the leading arm.
The member candidates in both areas fit well with the orbits, proper motions and $V\rm_{los}$ of trailing arm. Thus, we conclude the substructures detected in areas A and B belong to the Sgr stream. However, in the distance to the Sun, some member candidates apparently lie outside of the trailing arm of Sgr stream. The hypothesize some halo stars maybe mixed in the member candidates. To exclude these possible contaminants, only 106 stars whose distance to the Sun larger than 18 kpc were remained as the reliable members of those two substructures.

\begin{figure}
\plotone{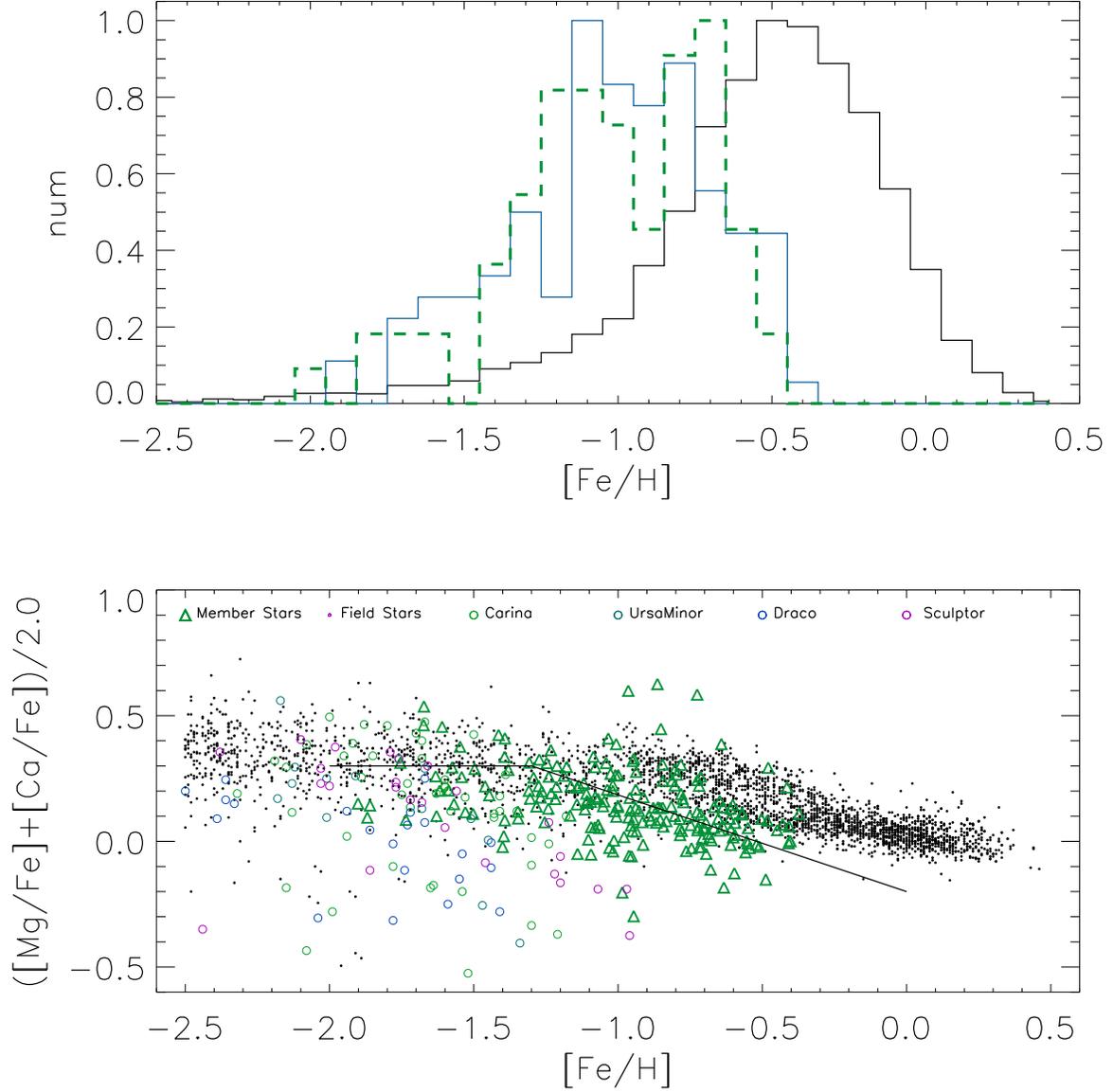}
\caption{Metallicity distribution and $\alpha$ abundance pattern for 106 member candidates. Top panel: Metalllicity distribution. Black solid line represents the normalized metallicity distribution for the whole sample. Blue solid line is metallicity distribution for member stars in area A while green dashed line for member stars in area B. Bottom panel: [$\alpha$/Fe] vs. [Fe/H]. Black points are Galactic field stars from SAGA database. Green diamonds represent the member candidates in our substructures. The colored open circles are some members of other dwarf galaxies from SAGA database.  \label{Fig5}}
\end{figure}

\begin{figure}
\plotone{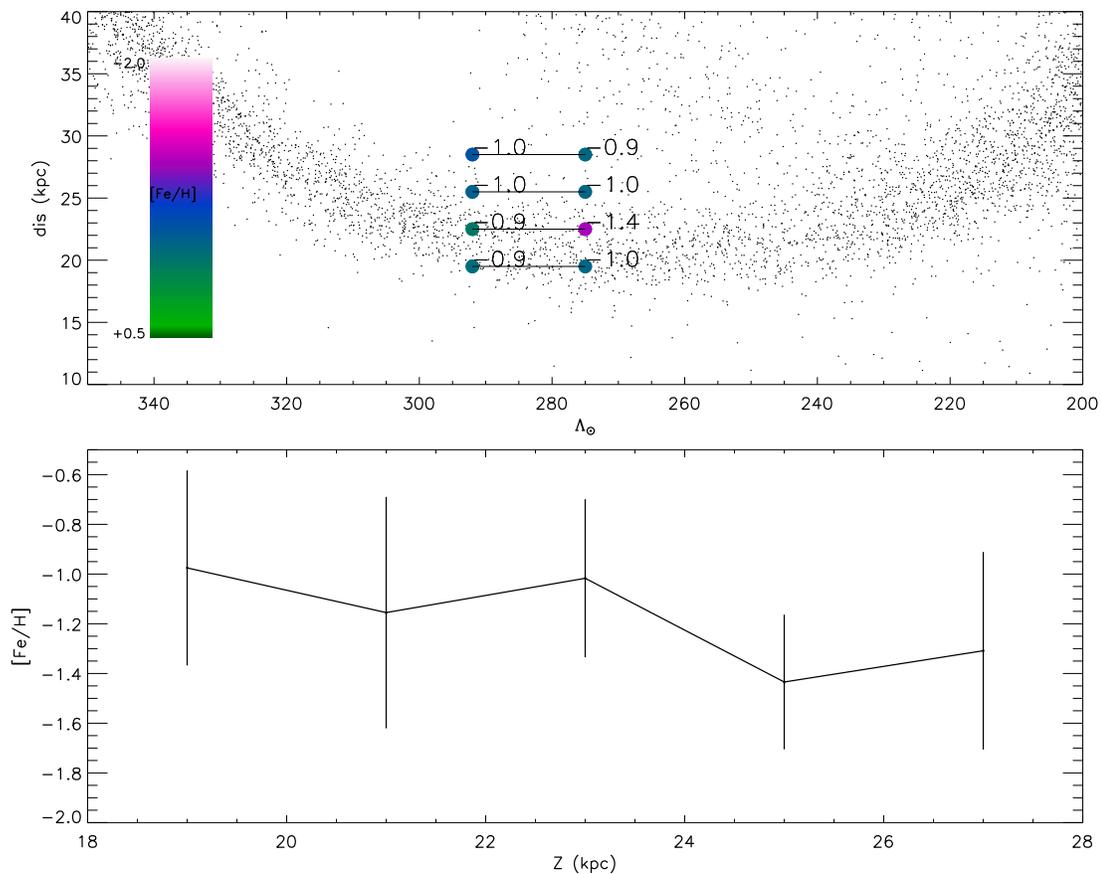}
\caption{ Top panel: [Fe/H] distribution in $\Lambda_{\odot}$ vs. {\bf distance to the Sun}. $\Lambda_{\odot}$ represents the coordinate of Sgr stream orbit. Filled points are four bins with different distance to the Sun in areas A and B. The color represents the mean metallicity in each bin.    Bottom panel: [Fe/H] vs. absolute Z. Black solid line connects the points with different Z value. The error bar of each points are from the [Fe/H] {\bf scatter} of each bin.   \label{Fig6}}
\end{figure}

\section{Metallicity and $\alpha$ abundance}
The chemical properties of stream candidates are very important metrics since the stars retain the primordial abundance of the original  molecular cloud. The abundance pattern of the member stars can be used to trace the formation and enrichment history of the progenitor cloud.
Monaco et al. (2007) presented radial velocities for 67 stars belonging to the Sgr stream and iron along with $\alpha$(Mg, Ca) abundances for 12 stars in the sample. The stream stars follow the same trend as Sgr main body stars in the [$\alpha$/Fe] vs. [Fe/H] plane. On average, the [Fe/H] of stars in this stream is metal poor compared to stars in the main body.

It has been shown that  the chemical abundance patterns of Sgr are quite unique. The
metal-rich Sgr stars ([Fe/H] $>$ -0.8) exhibit deficiencies in all chemical elements (expressed as [X/Fe]) relative to the MW
(see, e.g., McWilliam et al. 2013; Hasselquist et al. 2017). Recent work confirms these abundance patterns extend to the
Sgr tidal streams as well (Carlin et al. 2018). Carlin et al. 2018 analyzed [Fe/H] and [$\alpha$/Fe] of 42 member stars of the Sgr stream both in the leading and trailing arm. They found the average [Fe/H] for the stream is lower than that of the Sgr core and stars in the trailing and leading arms show systematic differences in [Fe/H].

Chou et al. (2010) presented high-resolution spectroscopic measurements of the abundances of the $\alpha$ element titanium (Ti) and s-process elements yttrium (Y) and lanthanum (La) for 59 candidate M giant members of the Sgr dSph + tidal tail system pre-selected on the basis of position and RV. The abundances of these stars show different pattern with Milky Way stars. The metallicity ranges from -1.4 to solar abundance. The overall [Ti/Fe], [Y/Fe], [La/Fe], and [La/Y] patterns with [Fe/H] of the Sgr stream plus Sgr core, for the most part resemble those seen in the Large Magellanic Cloud (LMC) and other dSphs, only shifted by $\Delta$[Fe/H] ~ +0.4 from the LMC and by ~+1 dex from the other dSphs.

Yang et al. (2019) found the trailing arm is on average more metal-rich than the leading arm. The $\alpha$ abundance of Sgr stars exhibit a similar trend to the Galactic halo stars at lower metallicity ([Fe/H] $<$ -1.0 dex), but extend to lower [$\alpha$/Fe] than Galactic disk stars at higher metallicity, which is close to the evolution pattern in dSphs.

\subsection{[Fe/H] and [$\alpha$/Fe] distribution}
Fig. 5 provides the chemical abundance pattern of $\alpha$ elements.
 The top panel shows the normalized metallicity distribution of stars in the whole sample (black solid line), members in area A (blue solid line) and members in area B (green dashed line). The metallicity distribution in area A and area B are very consistent, but very different from the whole sample. The [Fe/H] distribution of the stream centers on $\sim$ -0.9 and is lower than that of the whole sample. This indicates metal poor stars dominate the stream.  The bottom panel compares the $\alpha$ abundance patterns of stream member and  the halo field stars from the SAGA database (Suda et al. 2008) \footnote{http://sagadatabase.jp/}. Other dwarf galaxies such as Carina, Ursa Minor, Draco and Sculptor from SAGA database are also displayed with colored open circles. Fig. 5 shows the [$\alpha$/Fe] distribution among our stream candidates is very similar to that of Galactic field stars with [Fe/H] $<$ -1.5. However, among stars with [Fe/H] $>$ -1.0, [$\alpha$/Fe] is lower among stream candidates compared to field stars with the same [Fe/H]. The most likely explanation is that $\alpha$ abundances are mainly enriched by the contribution of supernova SN-II while iron peak elements are enriched by SN-Ia.   The evolution of  progenitor of stream would likely be slower than the Milky Way.  SN-Ia begin to contribute at lower [Fe/H], leading to the lower [$\alpha$/Fe] than Galactic stars.

 Most dSphs have different alpha-knee with the Milky Way which locates about [Fe/H] $\sim$ -1.0.  We estimate the alpha-knee of the Sgr stream is {\bf with [Fe/H] $\sim$ -1.1$\pm$ 0.2 (solid line)}. This value is similar to the results of Carretta et al. (2010) and de Boer et al. (2015). This alpha-knee is also consistent with the star formation history (SFH) of massive dwarf galaxies.



\subsection{Metallicity gradient}

Chou et al. (2007) investigated the variation of the metallicity distribution function (MDF) along the Sgr stream using M giants. The Sgr MDF was found to range from a median [Fe/H] $\sim$ -0.4 in the core to $\sim$ -1.1 dex over the Sgr leading arm length, representing $\sim$2.5-3.0 Gyr of dynamical (i.e. tidal stripping) age. Recently,  Hayes et al. (2020) identified 166 Sgr stream members observed by APOGEE that also have Gaia DR2 astrometry. They found a mild chemical gradient along the each arm.

Our stream candidates in areas A and B can be used to re-access the local metallicity gradient in the Sgr stream. Fig. 6 shows the [Fe/H] distribution among candidates in our sample. The top panel displays the [Fe/H] gradient along the orbit of the Sgr stream. The bottom panel shows the vertical [Fe/H] distribution.
In the top panel, $\Lambda_{\odot}$ represents the coordinate of the Sgr stream orbit. Filled points are four bins with different distance to the Sun in areas A and B. The color represents the mean metallicity in each bin. The distances range from 18 kpc to 30 kpc and the binsize is 3 kpc.  Thus, our data suggest a very small [Fe/H] gradient along the orbit, which is consistent with that of previous research. The bottom panel shows [Fe/H] gradient along the Z direction for stars in both A and B. The error bars are larger than the scatter. No evident metallicity gradients are found. Thus, along the stream orbit, we find a very small [Fe/H] gradient and no evident [Fe/H] gradient along the direction of vertical orbit among the member candidates in A and B. A more detailed investigation of the [Fe/H] gradients in the Sgr stream will require more data in much larger fields.

\section{Conclusion}
We identified two very prominent substructures in two areas of LCSSPASGC A and B. With the proper motions and $V\rm_{los}$, we identified 220 initial candidate stream members. Using the LM10 model as a reference,  two substructures were detected, which we propose are sections of the Sgr stream. As an additional filter on the preliminary sample, we excluded the candidates with distances to the Sun less than 18 kpc and obtained 106 likely member stars.

With the [Fe/H] and [$\alpha$/Fe] estimated from LAMOST spectra, the metallicity distribution of stars in stream and total sample were analyzed. We estimated the alpha-knee at [Fe/H] $\approx$ {\bf -1.1$\pm$0.2 dex}, which is consistent with that from the literature. Finally, the [Fe/H] gradients were studied.  [Fe/H] was found to vary slowly along the stream orbit between A and B.

The member stars in area A and B provide two small windows of Sgr stream.  Although large progress has been made in  the studies of the Sgr stream, more data in local fields along the orbit are needed to better characterize the entire stream and to constrain dynamical and chemical  models.

\acknowledgments
This study is supported by the National Natural Science Foundation of China under grant No. 11988101, 11973048, 11927804, 11890694, 11625313, 11733006  and National Key R\&D Program of China No. 2019YFA0405502. TDO acknowledges support from the U.S. National Science Foundation grant AST-1910396 to Embry-Riddle Aeronautical University.  This work is also supported by the Astronomical Big Data Joint Research Center, co-founded by the National Astronomical Observatories, Chinese Academy of Sciences and the Alibaba Cloud. Guoshoujing Telescope (the Large Sky Area Multi-Object Fiber Spectroscopic Telescope LAMOST) is a National Major Scientific Project built by the Chinese Academy of Sciences. Funding for the project has been provided by the National Development and Reform Commission. LAMOST is operated and managed by the National Astronomical Observatories, Chinese Academy of Sciences.


\clearpage

\end{document}